\documentclass[fleqn,usenatbib]{mnras}

\usepackage{newtxtext,newtxmath}

\usepackage[T1]{fontenc}

\DeclareRobustCommand{\VAN}[3]{#2}
\let\VANthebibliography\thebibliography
\def\thebibliography{\DeclareRobustCommand{\VAN}[3]{##3}\VANthebibliography}

\usepackage{graphicx}	
\usepackage{amsmath}	
\usepackage{booktabs}    
\usepackage{array, booktabs, makecell}

\title[Spectral Analysis of the LMXB XTE J1810--189]{Spectral Analysis of the LMXB XTE J1810--189 with \textit{NICER} Data}

\author[A. Manca et al.]{
A. Manca$^{1}$\thanks{E-mail: arianna.manca@dsf.unica.it},
A. Sanna$^{1}$,
A. Marino$^{4,5}$,
T. Di Salvo$^{2}$,
S. M. Mazzola$^{1}$,
A. Riggio$^{1,3}$,
\newauthor{N. Deiosso$^{1}$,
C. Cabras$^{1}$,
L. Burderi$^{1}$}
\\
$^{1}$ Universit\`a degli Studi di Cagliari, Dipartimento di Fisica, SP Monserrato-Sestu km 0.7, I-09042 Monserrato, Italy\\
$^{2}$ Universit\`a degli Studi di
  Palermo, Dipartimento di Fisica e Chimica, via Archirafi 36 - 90123 Palermo, Italy\\
$^{3}$ INAF/IASF Palermo, via Ugo La Malfa 153, I-90146 - Palermo, Italy\\
$^{4}$ Institute of Space Sciences (ICE, CSIC), Campus UAB, Carrer de Can Magrans s/n, E-08193 Barcelona, Spain \\
$^{5}$ Institut d'Estudis Espacials de Catalunya (IEEC), E-08034 Barcelona, Spain  \\
}

\date{Accepted XXX. Received YYY; in original form ZZZ}

\pubyear{2022}

\begin{document}
\label{firstpage}
\pagerange{\pageref{firstpage}--\pageref{lastpage}}
\maketitle

\begin{abstract}
XTE J1810--189 is a Low-Mass X-ray Binary transient system hosting a neutron star, which underwent a three-month-long outburst in 2020. In order to study its spectral evolution during this outburst, we analysed all the available observations performed by \textit{NICER}, in the 1--10~keV energy band. Firstly, we fitted the spectra with a thermal Comptonisation model. Our analysis revealed the lack of a significant direct emission from a black-body-like component, therefore we calculated the optical depth of the Comptonising region, deriving an upper limit of 4.5, which suggests the presence of a moderately thick corona. We also attempted to fit the spectrum with an alternative model, i.e. a cold Comptonised emission from a disc and a direct thermal component from the neutron star, finding a similarly good fit. The source did not enter a full high luminosity/soft state throughout the outburst, with a photon index ranging from $\sim$1.7 to $\sim$2.2, and an average unabsorbed flux in the 1--10~keV band of $\sim 3.6 \cdot 10^{-10}$ ~erg~cm$^{-2}$~s$^{-1}$. We searched for the presence of Fe K-shell emission lines in the range $\sim 6.4-7$~keV, significantly detecting a broad component only in a couple of observations. Finally, we conducted a time-resolved spectral analysis of the detected type-I X-ray burst, observed during the outburst, finding no evidence of a photospheric radius expansion. The type-I burst duration suggests a mix of H/He fuel.
\end{abstract}

\begin{keywords}
X-rays: binaries -- stars: neutron -- X-rays: individual: XTE J1810--189 -- accretion discs
\end{keywords}



\section{Introduction}

The binary system XTE J1810--189 has been classified as a Low-Mass X-ray Binary \citep[LMXB;][]{TOO2010}, a type of binary systems in which a compact object accretes matter from a companion star of mass $\lesssim$~1~M$_\odot$ \citep{lewin1980}. The compact object can either be a black hole (BH LMXB) or a neutron star (NS LMXB). The nature of the compact object determines different spectral \citep{mitsuda84} and temporal \citep{wijnands2001} properties of the source.

In LMXB systems, the central object accretes through Roche-lobe overflow, with the creation of an accretion disc around the object, which emits in the X-ray band. The emission is scattered and modified by a diffuse region of plasma around the central object (often called corona or Comptonising region), by means of inverse Compton scattering.

One of the features that allows to infer the nature of the compact object in these systems is the observation of type-I X-ray bursts, consisting of sudden nuclear burnings of piled-up accreted material on the surface of neutron stars \citep[see, e.g.,][for a review]{Galloway2021}. The spectra of type-I X-ray bursts can be well fitted with a black body model, radiated by the surface of the compact object \citep[][and ref. therein]{Marino2019, Kashyap2021}. For the brightest bursts, the emitted luminosity can reach the Eddington limit at the surface, causing the neutron star atmosphere to lift. As the atmosphere expands and its radius increases -- while the luminosity remains almost constant -- the effective temperature decreases ($L_b \propto R^2 T^4$). Type-I X-ray bursts with photospheric radius expansion (PRE) can therefore be recognised if the radius increases and the effective temperature decreases at the same time, while the flux remains constant. When the highest effective temperature is reached, the corresponding radius of the black body should represent the NS radius \citep{Galloway2021}. In the hypothesis that the luminosity remains equal to the Eddington luminosity during the burst, it is possible to give an estimate of the distance of the source from the observed flux \citep{Galloway2008}.

XTE J1810--189 is a transient source, which alternates between quiescent phases and periods of increased X-ray emission due to accretion on the central object. When in outburst, these transient systems typically evolve from a hard to a soft state \citep[see][for BH and NS systems respectively]{belloni2009, MunozDarias2014}. The hard states are characterised by X-ray emission at higher energies and lower luminosities in the soft X-ray band, with a predominance of Comptonisation processes in the surrounding, now expanded, region. The soft states instead occur when the mass accretion rate from the companion star is higher, leading to a more intense energy emission at low energy in the X-ray band, and to the cooling of the Comptonisation region possibly becoming smaller.

In NS LMXB systems, the soft states are decomposed into a soft and a hard component, in which the hard part is identified with a Comptonised component with the temperature of the corona, kT$_e$, of a few keV, while the soft part can be modelled either with a single temperature component (black body component from the neutron star), of temperature $\sim$~1.5~keV, or with a multi-colour disc (MCD) from the accretion disc, of lower temperature \citep[see, e.g.,][for a full review]{Barret2001}, or both. During the soft states, the optical depth of the Comptonising region can reach values of $\sim$~5--15, while the seed-photon temperature is compatible with the temperature of the soft component. The hard states are similarly decomposed into a hard and a soft component, but the optical depth of the Comptonising region reaches lower values, $\sim$~2--3. The soft component can be modelled as a single temperature black body or a multi-colour disc, when detectable. In both hard and soft states, an iron emission line can be detected at 6.4--6.7~keV, and sometimes also a weak reflection continuum at higher energies, called Compton hump. The iron line is thought to originate from the inner parts of the accretion disc, irradiated by the Comptonised emission and giving rise to recombination processes. The emitted energy is then altered by relativistic effects (Doppler shifts and gravitational redshifts), assuming a broad and asymmetric profile. It can be used to investigate the ionisation state and the geometry of the accretion flow \citep[see e.g.,][]{piraino12, DiSalvo2015, Degenaar2016, Ludlam2018}.

XTE J1810--189 was observed for the first time in 2008 by the Rossi X-ray Timing Explorer \citep[\textit{RXTE},][]{Tel1424} during a monitoring scan of the Galactic ridge region. Further \textit{RXTE} observations showed the presence of type-I X-ray bursts \citep{Tel1443}, confirming the nature of the compact object hosted in the system, while no burst oscillations have been found. A first upper limit to the distance was set to 11.5~kpc \citep{Tel1443} from the analysis of a type-I X-ray burst under the assumption that the peak flux was the Eddington flux. A stricter estimate on the distance was obtained by \citet{weng15} through a type-I X-ray burst analysis occurred during the 2008 outburst. Their results indicate a distance in the range 3.5--8.7~kpc. An infrared counterpart was detected thanks to the \textit{Chandra} localisation \citep{Tel1508, Tel1633} in 2008. In September 2020, new X-ray activity was observed by \textit{INTEGRAL} and subsequently by the Neutron star Interior Composition Explorer \citep{Tel14009}.

In this paper, we report the spectral analysis of the source using the data collected by \textit{NICER} during the 2020 outburst, with the aim to study the spectral evolution of the system. Since \textit{NICER} energy band covers lower values than \textit{RXTE}, we are able to study XTE J1810--189 for the first time at softer X-ray energy values.

\section{Data reduction and analysis}

\begin{figure}
 \includegraphics[width=\columnwidth]{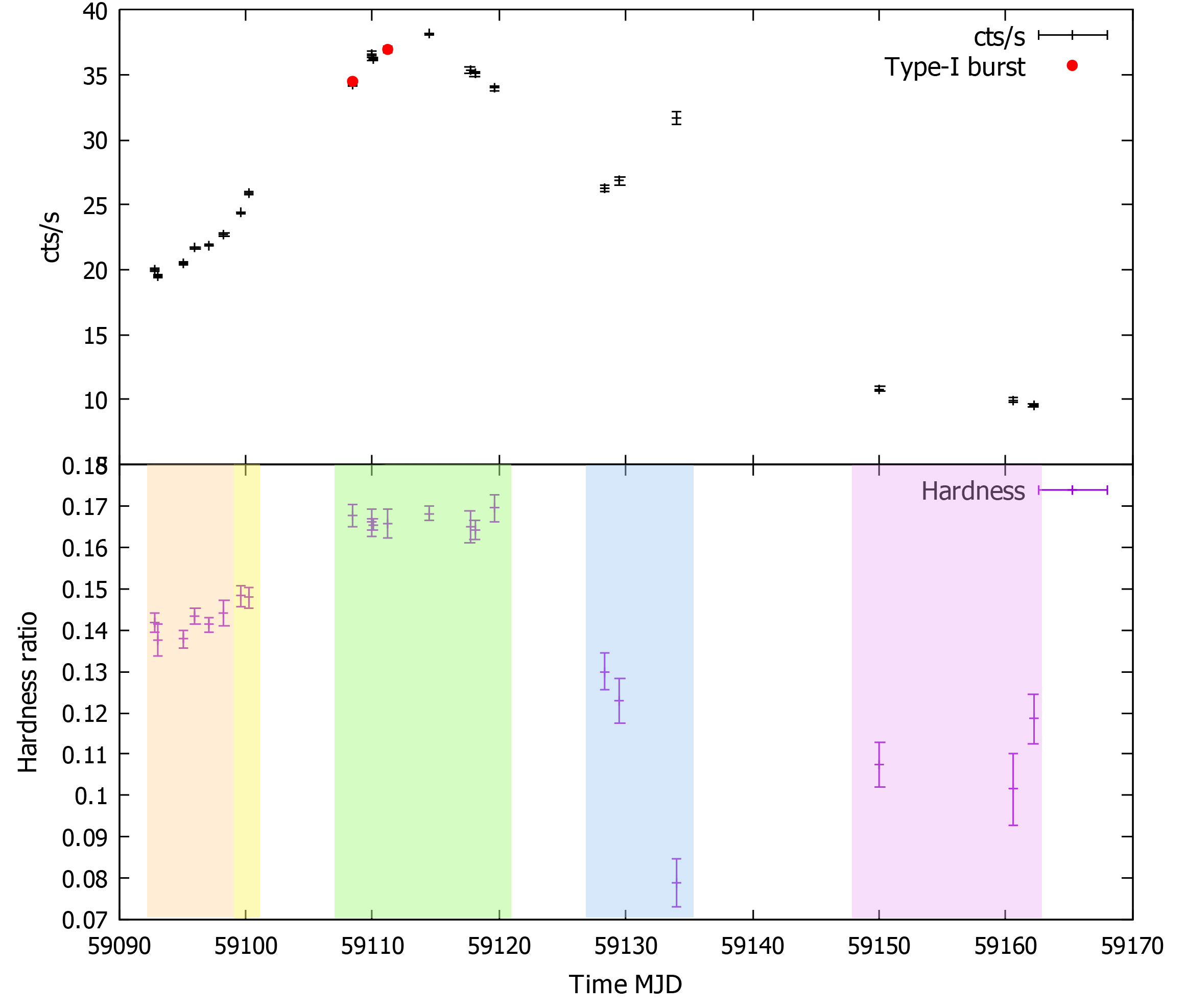}
 \caption{In the upper panel, we report the background-subtracted NICER light curve with 100 seconds binsize covering the entire 2020 outburst. Times are referred to J2000. The red dots represent type-I bursts. In the lower panel, we report the evolution of the hardness ratio with time. The hardness is defined as the ratio between the counts in the 6--10 and 0.6--3~keV energy bands. The coloured bands highlight the grouping done in the spectral analysis.}
 \label{fig:lc_hardness}
\end{figure}

\textit{NICER} observed the source from 2020, September 1 to 2020, November 13, for a total of 33 observations, covering the entire outburst (total exposure of 47793 seconds after data filtering, Figure \ref{fig:lc_hardness}). The outburst started at a rate of $\sim$~25 cts~s$^{-1}$, reached a peak of $\sim$~50~cts~s$^{-1}$ at MJD~59110, and decreased to $\sim$~10~cts~s$^{-1}$ towards the end. After filtering, only 23 observations have an exposure long enough to provide sufficient statistics for spectral analysis. Two observations (OBSID 3201750109 and OBSID 3201750112) show type-I X-ray bursts, marked in the Figure \ref{fig:lc_hardness} with red dots. Only one of the two bursts was fully observed by \textit{NICER}.

The reduction of \textit{NICER} data was performed using the pipeline \texttt{nicerl2} (\textsc{nicerdas} version 2020-04-23\_V007a), with default screening parameters: i) exclusion of time intervals in proximity of the South Atlantic Anomaly; ii) elevation angle of 30$^\circ$ over the Earth's limb; iii) minimum angle of 40$^\circ$ from the bright Earth limb; iv) maximum angular distance between the source direction and \textit{NICER} pointing direction of 0.015$^\circ$. The background modelling was done with the 3C50 model, gain epoch 2020 and the exclusion of the noisy FPM14 and FPM34 from the analysis. The spectra of the observations and the respective backgrounds were extracted using the tool \texttt{nibackgen3c50-v6f}. Observations OBSID09 and OBSID12 were also cleaned with the \texttt{xselect} tool, eliminating 140 seconds from the former observation and 110 seconds from the latter, in order to exclude the observed type-I X-ray bursts from the analysis. The spectral analysis was carried out in \textsc{xspec} 12.12.0 (\textsc{heasoft} package, version 6.29). The cleaned spectra were rebinned to contain at least 25 counts per channel.

Some peaks were visible below 1~keV, likely associated with the oxygen peak (0.56~keV), iron peak (0.71~keV) and neon peak (0.87~keV)\footnote{\url{https://heasarc.gsfc.nasa.gov/docs/nicer/data_analysis/workshops/RonR_spec_practices_20210511.pdf}}. In order to avoid the background-dominated part of the spectra and the instrumental noise, we performed the spectral analysis in the range 1--10~keV, which suits all the observations.

We also calculated the hardness ratio for each observation with the \texttt{xselect} tool, adopting the 0.6--3~keV energy range for the soft band, and the 6--10~keV energy range for the hard band.

\begin{table}
    \centering
    \caption{List of the observations with relative date in UTC format and exposure after filtering and type-I burst removal. Counts are the total counts. }
    \label{tab:OBSIDS}
    \begin{tabular}{lccr}
    	\hline
       OBSID & Date & Exposure [s] & Counts\\
       \hline
       3201750101 & 2020-08-31T20:22:21  & 3903.0  &  97774 \\
       3201750102 & 2020-09-01T01:01:02  & 1290.0  &  32074 \\
       3201750103 & 2020-09-03T01:08:20  & 3586.0  &  87114 \\
       3201750104 & 2020-09-04T00:17:19  & 4176.0  &  105645 \\
       3201750105 & 2020-09-05T02:37:21  & 5909.0  & 149798 \\
       3201750106 & 2020-09-06T06:37:00  & 1785.0  &  46743 \\
       3201750107 & 2020-09-07T13:41:00  & 2342.0  &  63659 \\
       3201750108 & 2020-09-08T05:11:20  & 2313.0  &  66444 \\
       3201750109 & 2020-09-16T09:43:32  & 1791.7  &  72622 \\
       3201750110 & 2020-09-17T22:54:32  & 1079.0  &  44272 \\
       3201750111 & 2020-09-18T00:39:20  & 7294.0  &  296877 \\
       3201750112 & 2020-09-19T05:53:52  & 1238.5  &  53461 \\
       3201750113 & 2020-09-22T11:21:09  & 3694.0  &  156144 \\
       3201750114 & 2020-09-25T16:54:18  & 869.0  &  35467 \\
       3201750115 & 2020-09-26T02:12:38  & 2168.0  &  89425 \\
       3201750116 & 2020-09-27T15:24:16  & 1264.0  &  52987 \\
       3201750118 & 2020-10-06T09:07:45  & 561.0  &  16561 \\
       3201750119 & 2020-10-07T12:45:00  & 340.0  &  10546 \\
       3201750120 & 2020-10-11T23:54:00  & 150.0  &  5292 \\
       3201750126 & 2020-10-28T01:41:20  & 935.0  &  13571 \\
       3201750132 & 2020-11-07T15:24:20  & 300.0  &  3467 \\
       3201750133 & 2020-11-09T04:38:12  & 805.0  &  9398 \\
       \hline
    \end{tabular}
\end{table}

\subsection{Persistent emission} \label{sec:continuum}
We fitted the the persistent emission with four different models:
\begin{enumerate}
\item \emph{model 1}: absorbed Comptonised emission (\texttt{TBabs*nthComp});
\item \emph{model 2}: absorbed Comptonised emission with the addition of a black body component. We investigated two variations of this model, one with the black body temperature free to vary and the second with the black body temperature fixed at 0.1~keV to account for an accretion disc emission. We rejected the first version since the addition of the black body was not significant (\texttt{TBabs*(bbodyrad+nthcomp)});
\item \emph{model 3}: absorbed black body component and power-law with low-energy roll-off (\texttt{TBabs*(bbodyrad+expabs*powerlaw)}).
\end{enumerate}

In the first model, we used the \texttt{nthComp} component \citep{zdziarski96, zycki99} to describe the thermally Comptonised continuum due to the presence of the electron corona. We set the \textit{inp\_type} parameter to 0, indicating a black body distribution for the seed photons. To take into account photoelectric absorption by the neutral matter in the ISM, we used the Tuebingen-Boulder model (\texttt{TBabs} in \textsc{xspec}) with the value of the equivalent hydrogen column density $n_H$ free to vary. The abundances were set to the values provided by \citet{wilms}. Since the lack of coverage above 10 keV does not allow us to properly constrain the electronic temperature kT$_e$, the latter was fixed to 30~keV for all the models, higher than \textit{NICER} upper energy limit and consistent with the typical values reported for similar sources \citep[e.g.,][]{Barret2001, Wang2017, Marino2019}. However, varying this parameter towards higher or lower values does not significantly affect the fit result, suggesting that the fit is insensitive to the electron temperature changes.

We attempted to fit the data with a second model that considers the presence of a soft component due to the neutron star emission. The model 2 did not result in a statistically significant improvement with respect to model 1. In fact, the application of an F-test, calculated at the 5\% significance level, between the two models showed that only 3 observations out of 22 (OBSID11, OBSID15, OBSID16) had a significant improvement with the addition of a black body component, although the best fit values of the photon index could not be constrained to physically reasonable values. We attempted to link the black body temperature of the \texttt{bbodyrad} component with the seed-photon temperature of the \texttt{nthComp} component, but the model did not give any significant improvement with respect to model 1. Therefore, we abandoned this version of model 2 and adopted a cold-disc variation. In order to explicitly search for a direct NS emission component, we kept the seed-photon temperature fixed at a very low value (kT$_{seed}$=~0.1~keV) with a disc-like distribution. Therefore, we hypothesised the presence of a cold disc that is not directly detectable but undergoes Comptonisation. The black body component, which in this case is significant for all the observations, can be identified with direct emission from the NS.

We could still see some residuals in the observations around 6.4~keV, therefore we searched for the presence of an iron emission line. We added a \texttt{gaussian} component to our models and constrained the centroid value to vary in the range 6.4--6.97~keV. In the case of model 1, the inclusion of the \texttt{gaussian} component was found to be statistically significant only in 8 out of 22 observations. For model 2, the addition of the gaussian component was significant for 10 observations out of 22. In order to improve the statistics, we decided to simultaneously fit the observations in five groups (OBSIDs 01--06; OBSIDs 07 and 08; OBSIDs 09--16; OBSIDs 18--20; OBSIDs 26, 32 and 33), in accordance with their hardness ratio (see Figure \ref{fig:lc_hardness}, lower panel) and temporal vicinity. For statistically non significant detections, we derived the upper limit on the equivalent width by fixing the values of the peak energy and line width to the values derived from the fit conducted with the same parameters free to vary.

To make a comparison with the previous study of XTE J1810--189, we fitted the \textit{NICER} data with a third model, composed by a black body component and a power-law in which the component \texttt{expabs} accounts for the low-energy roll-off, and the cut energy linked to the black body temperature. This model can be thought to be geometrically equivalent to our cold-disc version of model 2.

\begin{figure}
 \includegraphics[width=\columnwidth]{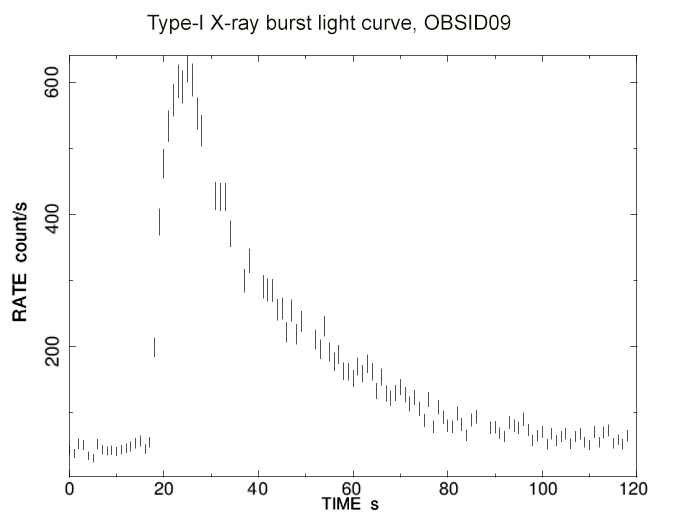}
 \caption{Light curve, of 1 second of binsize, of the type-I X-ray burst of OBSID09. Times are given with respect to MJD 59108.}
 \label{fig:burst}
\end{figure}

\subsection{Type-I X-ray bursts}
We identified two occurrences of type-I X-ray burst, in OBSIDs 09 and 12 respectively, but the latter was not observed in its entirety and it was not possible to fully observe the peak. Therefore, we analysed only the type-I X-ray burst of OBSID09, whose light curve is shown in Figure \ref{fig:burst}. We conducted a time-resolved spectral analysis, dividing the light curve in bins of $\sim$~2~seconds around the peak, and $\sim$~10 seconds on the tail, by using the \textsc{xselect} tool. Initially, we tried to fit the type-I X-ray burst without subtracting the persistent emission but fitting simultaneously the continuum and the burst instead. With such a strategy, we are taking into account the impact that the burst onset has on the accretion flow, usually able to amplify the persistent emission by a factor $f_a$ \citep{Worpel2013, Worpel2015}. We modelled the burst spectrum with an absorbed black body model (\texttt{TBabs*bbodyrad}) and used \texttt{nthcomp} (model 1) to describe the continuum. We therefore fitted each spectrum with the model: \texttt{TBabs*(bbodyrad+const*(nthcomp))}, with \texttt{const} as the factor $f_a$. When fitting the spectra with this model, we only allowed $f_a$ to change while we kept frozen the parameters of the continuum. Unfortunately, the poor statistics in the time-resolved spectra did not allow to fit both components and left $f_a$ totally unconstrained. Therefore, we adopted the persistent emission as background for the observation \citep[see, e.g.,][]{Galloway2008}. We extracted 10 seconds of the persistent emission before of the burst, to use as background. We rebinned all spectra in order to have at least 20 counts per bin and we modelled the burst spectrum with an absorbed black body model (\texttt{TBabs*bbodyrad}).

We ran the initial fit leaving the hydrogen column density free to vary, but we noticed a correlation between the n$_H$ and the black body normalisation parameters. Therefore, we kept n$_H$ fixed to the value obtained from the persistent analysis with model 1 (see Table \ref{tab:model1}) for OBSID09 in the final analysis of the burst.

\subsection{Timing analysis}
\begin{figure}
 \includegraphics[width=\columnwidth]{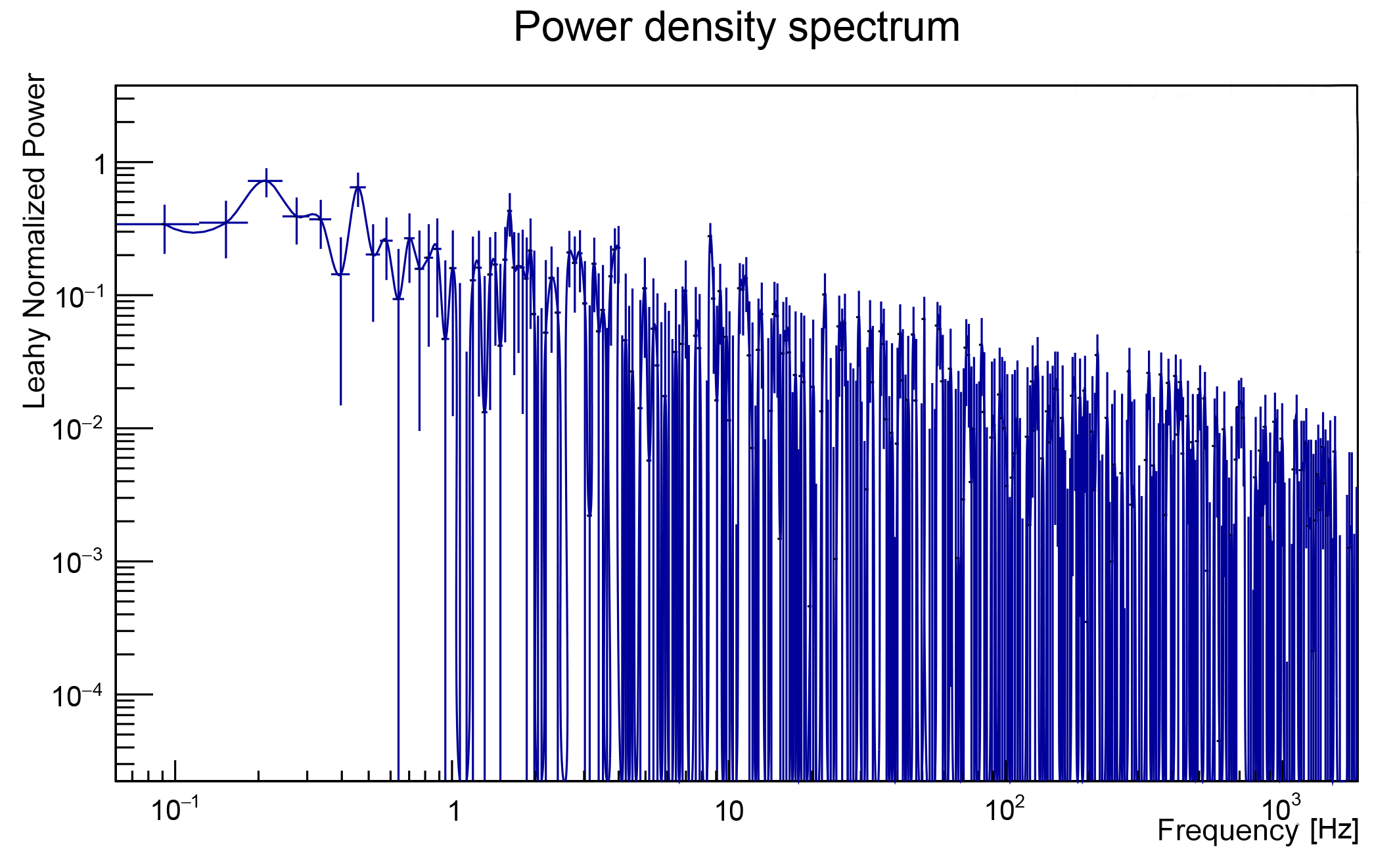}
 \caption{Power density spectra of the observation OBSID01, with bin size 250 $\mu s$ and Nyquist frequency 4096~Hz.}
 \label{fig:PDS}
\end{figure}

In order to conduct a timing analysis, we barycentered the event files with respect to the barycentre of the Solar System with the \texttt{barycorr} routine. We derived the Leahy-normalised power density spectra \citep[PDS,][]{Leahy1983} with the \texttt{powspec} command, setting a Nyquist frequency of 4096~Hz. We considered time intervals of 16~s of exposure for each observation, and averaged the resulting PDS accordingly. We estimated the Poissonian white noise with an unweighted fit of the PDS beyond 500~Hz with a constant, and subtracted it from the entire range. Figure \ref{fig:PDS} shows the PDS for OBSID01, as a reference. From each observation, we derived the fractional root mean square (RMS) up to 100~Hz, following the method described in \citet{BelloniHasinger}, and studied its evolution with time. To take into account the decreasing number of counts in the last observations, we merged the observations from OBSID18 to OBSID33 in a single PDS.

\section{Results}
Figure \ref{fig:lc_hardness} reports the evolution of the hardness ratio during the outburst. We notice an initial hardening of the source as it approaches the peak of the outburst, followed by a significant softening towards the end of the outburst. This behaviour is supported also by the spectral analysis of the continuum, reported in the following section.

\subsection{Persistent emission}\label{ref:results}
\begin{figure}
 \includegraphics[width=\columnwidth]{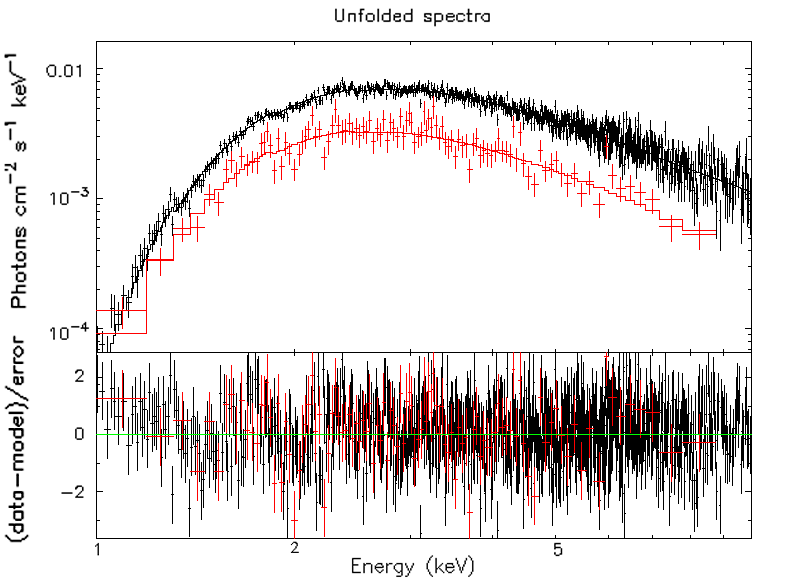}
 \caption{Unfolded spectra fitted with model 1. The black data belong to the OBSID01, while the red data are OBSID32.}
 \label{fig:ufspec}
\end{figure}

The results of the fit with model 1 show a variation of the photon index between 1.67$_{-0.06}^{+0.07}$ and 2.19$_{-0.27}^{+0.49}$ (errors at 90\% c.l.), during the outburst evolution, reaching the highest value in OBSID32 and never fully entering the high/soft state (same behaviour reported by \citealt{weng15}). The unabsorbed flux in the 1--10~keV energy range is around $10^{-10}$~erg~cm$^{-2}$~s$^{-1}$, which, with the upper limit on the distance of 11.5 kpc obtained by \citet{Tel1443}, leads to a luminosity upper limit of $\sim 5 \cdot10^{36}$~erg~s$^{-1}$ on average, showing that the source did not reach very high luminosities in this phase (i.e., of the order of 1\% of the Eddington luminosity for a NS of $1.4~M_\odot$). The obtained results are in satisfying agreement with the data, with reduced $\chi^2 \sim 1$ for all observations (Table \ref{tab:model1}). Figure \ref{fig:ufspec} shows the comparison between the spectra of the first observation and of the one with highest photon index value (OBSID32).

The seed photons temperature for the Comptonised spectrum is estimated between 0.47$_{-0.13}^{+0.11}$ and  0.79$\pm 0.05$~keV, consistent with the temperatures of the accretion disc reported in the literature for NS LMXBs \citep[see, e.g.,][]{gambino2019, Ludlam2020}. Since this model lacks a direct black body component, it is not obvious if the observed seed photons source for the Comptonisation component is the disc or the neutron star. In order to search for a direct, non Comptonised component from the neutron star, we added a black body component to model 1, obtaining model 2. The temperature of this second component, when left free to vary, returned best fit values around 2~keV for several observations, but the photon index was not able to converge to physically reasonable values.

\begin{table*}
    \centering
    \caption{Fit results for model 1 (\texttt{TBabs*nthcomp}) calculated in the 1--10~keV energy range. The reported flux is the unabsorbed flux in units of 10$^{-10}$~erg~cm$^{-2}~s^{-1}$. Errors are in the 90\% confidence level.}
    \label{tab:model1}
\renewcommand{\arraystretch}{1.3}
     \begin{tabular}{lcccccc}
     	\hline
        OBSID & n$_H [10^{22}$~cm$^{-2}]$ & Photon Index & kT$_{seed}$ [keV] & Flux [$10^{-10}$~erg~cm$^{-2}$~s$^{-1}$] & $\chi^2$/dof & $\tau$\\ \hline
       3201750101 & 3.67$_{-0.11}^{+0.12}$  &  1.95$_{-0.06}^{+0.06}$ &  0.63$_{-0.04}^{+0.04}$ &  2.99$_{-0.05}^{+0.06}$ & 709.88/691 & <3.87\\
       3201750102 & 3.72$_{-0.19}^{+0.21}$  &  1.99$_{-0.10}^{+0.13}$ &  0.63$_{-0.07}^{+0.07}$ &  2.89$_{-0.09}^{+0.10}$ & 542.21/497 & <3.79\\
       3201750103 & 3.68$_{-0.11}^{+0.12}$  &  1.97$_{-0.06}^{+0.06}$ &  0.62$_{-0.04}^{+0.04}$ &  2.96$_{-0.05}^{+0.06}$ & 697.98/652 & <3.83\\
       3201750104 & 3.54$_{-0.10}^{+0.10}$  &  1.98$_{-0.06}^{+0.07}$ &  0.67$_{-0.03}^{+0.03}$ &  3.08$_{-0.05}^{+0.05}$ & 720.65/690 & <3.81\\
       3201750105 & 3.54$_{-0.08}^{+0.08}$  &  2.02$_{-0.05}^{+0.06}$ &  0.68$_{-0.03}^{+0.03}$ &  3.07$_{-0.04}^{+0.04}$ & 699.51/731 & <3.71\\
       3201750106 & 3.75$_{-0.15}^{+0.17}$  &  1.93$_{-0.08}^{+0.09}$ &  0.62$_{-0.06}^{+0.05}$ &  3.28$_{-0.08}^{+0.09}$ & 534.38/565 & <3.90\\
       3201750107 & 3.52$_{-0.12}^{+0.12}$  &  2.01$_{-0.08}^{+0.09}$ &  0.70$_{-0.04}^{+0.04}$ &  3.38$_{-0.03}^{+0.03}$ & 681.10/621 & <3.75\\
       3201750108 & 3.56$_{-0.12}^{+0.12}$  &  2.03$_{-0.08}^{+0.09}$ &  0.70$_{-0.04}^{+0.04}$ &  3.58$_{-0.07}^{+0.07}$ & 616.04/615 &<3.71\\
       3201750109 & 3.42$_{-0.12}^{+0.12}$  &  2.00$_{-0.09}^{+0.11}$ &  0.79$_{-0.05}^{+0.05}$ &  4.99$_{-0.09}^{+0.09}$ & 664.65/634 & <3.76\\
       3201750110 & 3.43$_{-0.14}^{+0.15}$  &  1.95$_{-0.10}^{+0.12}$ &  0.75$_{-0.06}^{+0.06}$ &  5.19$_{-0.11}^{+0.12}$ & 599.90/563 & <3.87\\
       3201750111 & 3.46$_{-0.06}^{+0.06}$  &  1.87$_{-0.04}^{+0.04}$ &  0.72$_{-0.02}^{+0.02}$ &  5.24$_{-0.05}^{+0.05}$ & 902.20/811 & <4.03\\
       3201750112 & 3.37$_{-0.14}^{+0.14}$  &  1.95$_{-0.10}^{+0.13}$ &  0.77$_{-0.06}^{+0.06}$ &  5.31$_{-0.12}^{+0.12}$ & 556.75/572 & <3.86\\
       3201750113 & 3.42$_{-0.08}^{+0.08}$  &  1.78$_{-0.05}^{+0.05}$ &  0.69$_{-0.03}^{+0.03}$ &  5.48$_{-0.07}^{+0.07}$ & 712.44/741 & <4.24\\
       3201750114 & 3.52$_{-0.17}^{+0.18}$  &  1.79$_{-0.09}^{+0.11}$ &  0.67$_{-0.07}^{+0.07}$ &  5.20$_{-0.11}^{+0.12}$ & 519.32/527 & <4.24\\
       3201750115 & 3.38$_{-0.10}^{+0.11}$  &  1.83$_{-0.06}^{+0.07}$ &  0.70$_{-0.04}^{+0.04}$ &  5.14$_{-0.08}^{+0.07}$ & 724.93/673 & <4.13\\
       3201750116 & 3.56$_{-0.16}^{+0.18}$  &  1.67$_{-0.06}^{+0.07}$ &  0.58$_{-0.06}^{+0.06}$ &  5.36$_{-0.13}^{+0.15}$ & 612.88/595 & <4.55\\
       3201750118 & 3.25$_{-0.23}^{+0.28}$  &  1.80$_{-0.10}^{+0.13}$ &  0.59$_{-0.10}^{+0.09}$ &  3.50$_{-0.14}^{+0.16}$ & 422.73/395 & <4.21\\
       3201750119 & 3.23$_{-0.30}^{+0.49}$  &  1.82$_{-0.12}^{+0.18}$ &  0.55$_{-0.17}^{+0.12}$ &  3.53$_{-0.19}^{+0.31}$ & 279.78/286 & <4.16\\
       3201750120 & 3.81$_{-0.89}^{+0.24}$  &  1.85$_{-0.12}^{+0.18}$ &  $< 0.36$ &  4.47$_{-0.77}^{+0.15}$ & 181.69/151 & <4.09\\
       3201750126 & 3.38$_{-0.28}^{+0.33}$  &  2.00$_{-0.14}^{+0.20}$ &  0.59$_{-0.10}^{+0.09}$ &  1.53$_{-0.08}^{+0.09}$ & 341.17/312 & <3.76\\
       3201750132 & 3.51$_{-0.48}^{+0.63}$  &  2.19$_{-0.27}^{+0.49}$ &  0.61$_{-0.17}^{+0.16}$ &  1.26$_{-0.11}^{+0.15}$ & 127.96/107 & <3.42\\
       3201750133 & 3.94$_{-0.39}^{+0.52}$  &  1.96$_{-0.11}^{+0.14}$ &  0.47$_{-0.13}^{+0.11}$ &  1.35$_{-0.09}^{+0.14}$ & 217.60/247 & <3.85\\
       \hline
    \end{tabular}
\end{table*}

To explain the lack of a direct black body component, we derived the optical thickness $\tau$ from the values of the photon index (for the corresponding electron temperature, even if we fixed its value to 30~keV), following the formula given by \citet{zdziarski96}. The optical depth can be calculated in the assumption of a spherical geometry and uniform density. In that case, the energy photon index is equal to:
\begin{equation}
    \alpha = \bigg [\frac{9}{4}+\frac{1}{(kT_e/m_e c^2) \tau (1+\tau/3)}\bigg ]^{\frac{1}{2}} -\frac{3}{2},
    \label{eq:alpha}
\end{equation}

\noindent where kT$_e$ is the temperature of the corona. Since our value of the electron temperature was fixed at an arbitrary value, we cannot exactly measure the optical depth in our system. We are only able to determine upper limits on the optical depth, at 3$\sigma$ confidence level, by setting the electron temperature to the highest value (10~keV) in the range of our analysis. The estimated values of $\tau$ for each observation are reported in the last column of Table \ref{tab:model1}. The upper limits for the optical thickness vary in the range 3.42--4.55, with the highest value in correspondence of OBSID16. 

In the case of model 2, the photon index ranges from 1.23$^{+0.39}_{-0.21}$ to 2.19$^{+1.06}_{-0.51}$ in correspondence with OBSID20, after steadily increasing from the beginning of the outburst. The black body temperature varies from 0.70$^{+0.10}_{-0.13}$ to 1.44$^{+0.45}_{-0.52}$~keV (with the exception of OBSID20, for which the value of $kT_{BB}$ is completely unconstrained). Table \ref{tab:model4} reports the best fit values for the cold-disc version of model 2. We can use the normalisation of the black body component to give an estimate of the size of the emitting region, as derived from the relation:

\begin{equation}
\label{eq:norm}
    {\rm norm} = \frac{R_{{\rm km}}^2}{D_{10~{\rm kpc}}^2},
\end{equation}

\noindent where we assumed the upper limit of 11.5~kpc as a measure for the distance. We obtain an average upper limit of 4~km of radius, without any corrections applied. The small dimensions of the emitting region suggests a possible association with a hot spot on the surface of the neutron star \citep[see, e.g.,][]{Pintore2018}.

\begin{figure*}
 \includegraphics[width=0.9\textwidth]{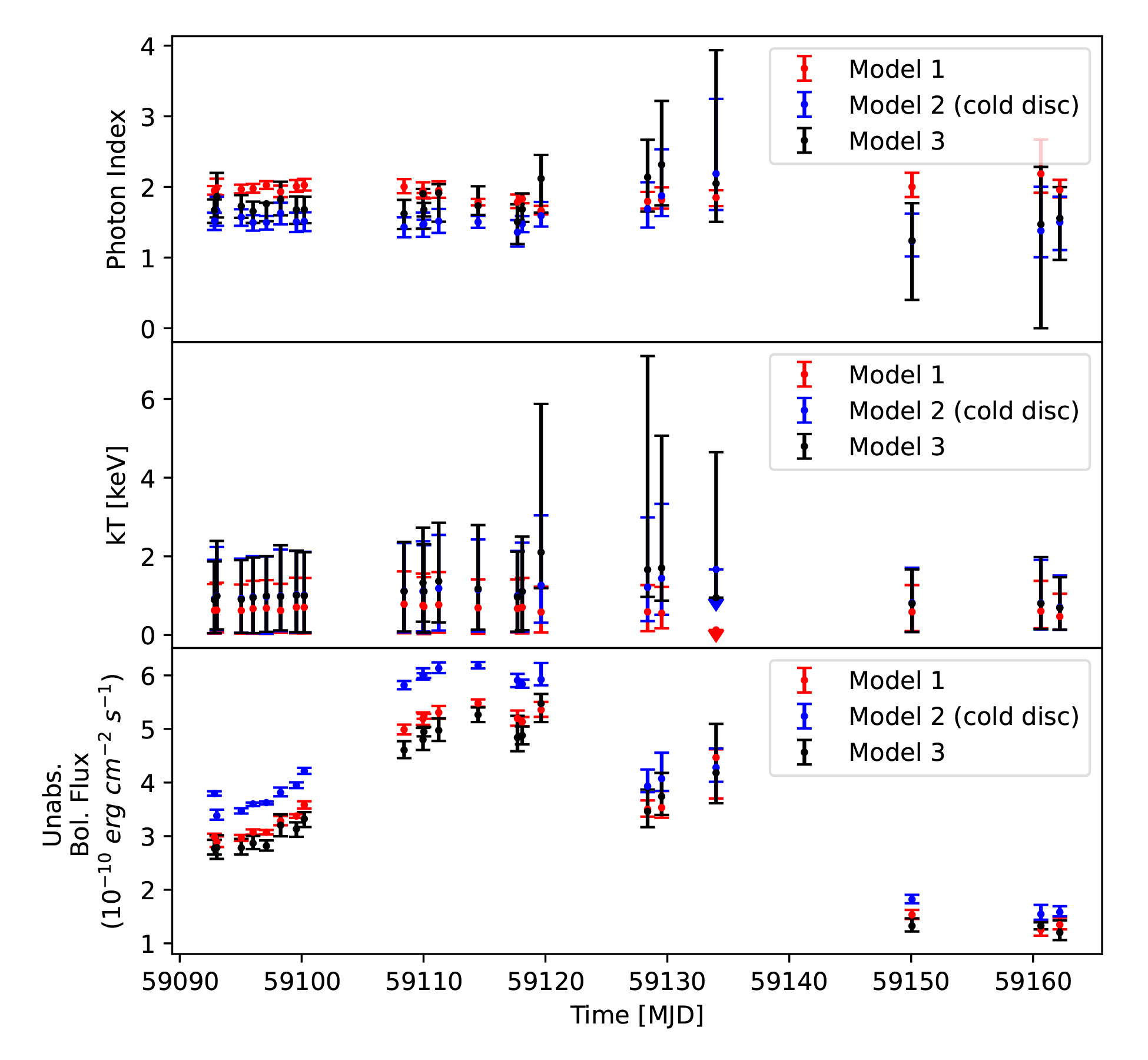}
 \caption{Temporal evolution of the photon index, the black body temperature and the unabsorbed flux in the 1--10~keV energy range, for each observation. The red series indicates model 1 (\texttt{TBabs*(nthComp)}); the blue series indicates model 2 in the cold-disc version (\texttt{TBabs*(bbodyrad+nthComp)}); the black series indicates model 3  (\texttt{TBabs*(bbodyrad+expabs*powerlaw)}). The error bars are at a 90\% confidence level.}
 \label{fig:parameters}
\end{figure*}

\begin{table*}
	\centering
	\caption{Fit results for model 2 in the cold-disc version (\texttt{TBabs*(bbodyrad+nthComp)}), where the seed photons temperature is fixed at 0.1~keV. The reported flux is the unabsorbed flux in the energy range 1--10~keV, in units of $10^{-10}$~erg~cm$^{-2}$~s$^{-1}$. Errors are at 90\% confidence level.}
	\label{tab:model4}
\renewcommand{\arraystretch}{1.3}
	\begin{tabular}{lcccccr}
		\hline
		OBSID & n$_H$ [10$^{22}$~cm$^{-2}]$ & Photon Index & kT$_{BB}$ [keV] & norm$_{BB}$ & Flux [$10^{-10}$~erg~cm$^{-2}$~s$^{-1}$] & $\chi^2$/dof\\
		\hline
       3201750101 & 4.2$_{-0.2}^{+0.2}$  &  1.52$_{-0.13}^{+0.12}$ &  0.92$_{-0.05}^{+0.07}$ & 9.0$_{-3.2}^{+3.4}$ &  3.80$_{-0.04}^{+0.04}$ & 717.14/690\\
       3201750102 & 4.5$_{-0.4}^{+0.3}$  &  1.67$_{-0.22}^{+0.20}$ &  0.10$_{-0.12}^{+0.24}$ & 5.6$_{-3.4}^{+5.7}$ & 3.38$_{-0.08}^{+0.11}$ & 541.78/496\\
       3201750103 & 4.3$_{-0.2}^{+0.2}$  &  1.57$_{-0.12}^{+0.11}$ &  0.93$_{-0.05}^{+0.08}$ & 8.2$_{-3.1}^{+3.5}$ & 3.47$_{-0.05}^{+0.05}$ & 703.82/651\\
       3201750104 & 4.1$_{-0.2}^{+0.2}$  &  1.50$_{-0.12}^{+0.10}$ &  0.97$_{-0.04}^{+0.06}$ & 8.4$_{-2.5}^{+2.7}$ & 3.60$_{-0.04}^{+0.03}$ & 723.28/689\\
       3201750105 & 4.1$_{-0.2}^{+0.3}$  &  1.50$_{-0.10}^{+0.10}$ &  0.97$_{-0.04}^{+0.05}$ & 8.9$_{-2.1}^{+2.3}$ & 3.63$_{-0.03}^{+0.02}$ & 714.49/730\\
       3201750106 & 4.5$_{-0.3}^{+0.3}$  &  1.63$_{-0.16}^{+0.15}$ &  0.99$_{-0.10}^{+0.19}$ & 6.1$_{-3.4}^{+4.8}$ & 3.81$_{-0.07}^{+0.09}$ & 534.97/564\\
       3201750107 & 4.2$_{-0.2}^{+0.2}$  &  1.51$_{-0.14}^{+0.13}$ &  1.02$_{-0.06}^{+0.09}$ & 7.9$_{-2.8}^{+3.3}$ & 3.95$_{-0.05}^{+0.06}$ & 687.66/620\\
       3201750108 & 4.2$_{-0.3}^{+0.3}$  &  1.51$_{-0.14}^{+0.13}$ &  1.01$_{-0.06}^{+0.09}$ & 8.9$_{-3.0}^{+3.5}$ & 4.22$_{-0.05}^{+0.06}$ & 625.03/614\\
       3201750109 & 4.1$_{-0.2}^{+0.2}$  &  1.43$_{-0.14}^{+0.14}$ &  1.12$_{-0.07}^{+0.10}$ & 9.6$_{-3.0}^{+3.6}$ & 5.82$_{-0.07}^{+0.08}$ & 670.06/633\\
       3201750110 & 4.1$_{-0.3}^{+0.3}$  &  1.47$_{-0.17}^{+0.17}$ &  1.11$_{-0.09}^{+0.16}$ & 9.1$_{-3.9}^{+5.0}$ & 6.01$_{-0.09}^{+0.12}$ & 609.88/562\\
       3201750111 & 4.2$_{-0.1}^{+0.1}$  &  1.48$_{-0.06}^{+0.06}$ &  1.11$_{-0.05}^{+0.06}$ & 8.1$_{-1.7}^{+1.9}$ & 6.00$_{-0.04}^{+0.05}$ & 925.53/810\\
       3201750112 & 4.1$_{-0.3}^{+0.3}$  &  1.52$_{-0.17}^{+0.17}$ &  1.19$_{-0.12}^{+0.17}$ & 7.5$_{-2.9}^{+4.3}$ & 6.13$_{-0.09}^{+0.11}$ & 556.87/571\\
       3201750113 & 4.2$_{-0.1}^{+0.1}$  &  1.50$_{-0.08}^{+0.09}$ &  1.15$_{-0.09}^{+0.13}$ & 6.1$_{-2.1}^{+2.6}$ & 6.18$_{-0.05}^{+0.07}$ & 727.29/740\\
       3201750114 & 4.0$_{-0.3}^{+0.3}$  &  1.36$_{-0.20}^{+0.17}$ &  1.00$_{-0.08}^{+0.13}$ & 11.7$_{-6.9}^{+3.9}$ & 5.91$_{-0.13}^{+0.12}$ & 518.66/526\\
       3201750115 & 4.1$_{-0.2}^{+0.2}$  &  1.47$_{-0.11}^{+0.11}$ &  1.11$_{-0.09}^{+0.13}$ & 7.5$_{-2.9}^{+3.7}$ & 5.84$_{-0.07}^{+0.08}$ & 735.25/672\\
       3201750116 & 4.3$_{-0.3}^{+0.1}$  &  1.59$_{-0.15}^{+0.19}$ &  1.26$_{-0.31}^{+0.52}$ & 2.4$_{-1.6}^{+5.8}$ & 5.92$_{-0.11}^{+0.31}$ & 612.30/594\\
       3201750118 & 4.0$_{-0.3}^{+0.8}$  &  1.69$_{-0.26}^{+0.38}$ &  1.21$_{-0.35}^{+0.57}$ & 2.3$_{-1.7}^{+8.0}$ & 3.93$_{-0.11}^{+0.31}$ & 423.14/394\\
       3201750119 & 4.1$_{-0.4}^{+0.5}$  &  1.87$_{-0.29}^{+0.66}$ &  1.44$_{-0.52}^{+0.45}$ & 1.6$_{-1.1}^{+4.9}$ & 4.07$_{-0.23}^{+0.49}$ & 274.30/285\\
       3201750120 & 4.0$_{-0.3}^{+0.8}$  &  2.19$_{-0.51}^{+1.06}$ &  1.67$_{-0.92}^{+4.41}$ & $< 3.1$ & 4.28$_{-0.27}^{+0.35}$ & 180.58/150\\
       3201750126 & 3.5$_{-0.4}^{+0.5}$  &  1.23$_{-0.22}^{+0.39}$ &  0.82$_{-0.08}^{+0.07}$ & 11.1$_{-5.9}^{+4.3}$ & 1.82$_{-0.07}^{+0.08}$ & 338.96/311\\
       3201750132 & 3.8$_{-0.7}^{+1.0}$  &  1.38$_{-0.38}^{+0.62}$ &  0.81$_{-0.14}^{+0.29}$ & 9.6$_{-8.3}^{+8.0}$ & 1.55$_{-0.10}^{+0.17}$ & 127.36/106\\
       3201750133 & 4.2$_{-0.5}^{+0.6}$  &  1.50$_{-0.39}^{+0.36}$ &  0.71$_{-0.13}^{+0.10}$ & 11.9$_{-9.0}^{+8.9}$ & 1.58$_{-0.08}^{+0.11}$ & 216.40/246\\
		\hline
	\end{tabular}
\end{table*}

Previous analyses of this source during the 2008 outburst \citep{weng15} revealed a softening of the spectra throughout the \textit{RXTE} observations. The range of the photon index values reported here is similar to the one of \citet{weng15}, even if the models are different. The authors used a model composed by a black body and a power-law component (without the energy cut-off), which we reproduced in model 3. In our model, the black body temperature, kT$_{BB}$, varies between 0.69$_{-0.13}^{+0.09}$ and 2.10$_{-1.19}^{+1.67}$~keV, while the photon index ranges between 1.24$_{-0.84}^{+0.53}$ and 2.32$_{-0.58}^{+0.90}$, reaching lower values than model 1 for most of the observations. The photon index slowly evolves from 1.67$_{-0.18}^{+0.15}$ in OBSID01 to its highest value of 2.32$_{-0.58}^{+0.90}$ in OBSID19. Table \ref{tab:model3} reports the best fit parameters for model 3. Following the same procedure as before, we find the size of the emitting region to be $\sim 3$~km. However, it should be noted that the normalisation of the black body component of model 3 is not well constrained.

Figure \ref{fig:parameters} compares the temporal evolution of the photon index (in red, the parameter from the \texttt{nthComp} component of model 1;  in blue, the photon index relative to the cold disc model (model 2); in black, the index of the power-law of model 3; and the black body temperature (temperature of the seed photons for the model 1). The figure shows a similar photon index evolution for models 2 and 3, which both imply a cold disc, contrary to model 1 in which the disc is relatively hot. The two different hypotheses lead to a slight difference in values between the two interpretations (model 1 for the hot disc hypothesis, models 2 and 3 for the cold disc hypothesis). In the case of the cold disc models, the photon index reaches its highest value earlier in time with respect to model 1. Moreover, the parameters of the models 3 (and of the observation OBSID20 with the model 2) have much larger uncertainties. The errors are particularly large in correspondence of OBSID20, which is also one of the observations with lower counts. The best fit value of the temperature parameter of model 1 is $\sim 1 \sigma$ lower than the values of the other observations, and the low statistics of OBSID20 may be a reason for this poor estimation. In model 2, the highest temperature (corresponding precisely to OBSID20) is 1.67$_{-0.92}^{+4.41}$~keV, still loosely constrained. It is worth to remind that the reported temperature for model 1 is the seed-photon temperature from a component that we cannot directly see, while for models 2 and 3 the reported temperature is the direct black body emission. Figure \ref{fig:parameters} reports the temporal evolution of the unabsorbed flux in the range 1--10~keV for the different models and, while the evolution follows the same track, in the case of model 2 the unabsorbed flux reaches higher values, due to a higher estimation of the hydrogen column density in this model. Because of the similar behaviour of models 2 and 3 and since model 2 is an alternative, more sophisticated formulation of model 3, we reduce the rest of the analysis to models 1 and 2.

\begin{table*}
	\centering
	\caption{Fit results for model 3 (\texttt{TBabs*(bbodyrad+expabs*powerlaw)}). The reported flux is the unabsorbed flux in the energy range 1--10~keV, in units of $10^{-10}$~erg~cm$^{-2}$~s$^{-1}$. Errors are at 90\% confidence level.}
	\label{tab:model3}
\renewcommand{\arraystretch}{1.3}
	\begin{tabular}{lcccccr}
		\hline
		OBSID & n$_H$ [10$^{22}$~cm$^{-2}]$ & Photon Index & kT$_{BB}$ [keV] & norm$_{BB}$ & Flux [$10^{-10}$~erg~cm$^{-2}$~s$^{-1}$] & $\chi^2$/dof\\
		\hline
        3201750101 & 3.99$_{-0.22}^{+0.19}$  &  1.67$_{-0.18}^{+0.15}$ &  0.89$_{-0.05}^{+0.08}$ & 8.7$_{-3.8}^{+4.4}$ &  2.77$_{-0.12}^{+0.16}$ & 715.49/690\\
       3201750102 & 4.20$_{-0.35}^{+0.27}$  &  1.86$_{-0.29}^{+0.34}$ &  0.99$_{-0.14}^{+0.42}$ & 4.7$_{-4.7}^{+7.0}$ & 2.79$_{-0.21}^{+0.23}$ & 541.65/496\\
       3201750103 & 4.05$_{-0.20}^{+0.19}$  &  1.73$_{-0.17}^{+0.16}$ &  0.90$_{-0.06}^{+0.10}$ & 7.9$_{-3.8}^{+4.3}$ & 2.78$_{-0.13}^{+0.17}$ & 702.03/651\\
       3201750104 & 3.88$_{-0.19}^{+0.17}$  &  1.65$_{-0.16}^{+0.14}$ &  0.95$_{-0.05}^{+0.07}$ & 7.9$_{-2.9}^{+3.5}$ & 2.87$_{-0.11}^{+0.14}$ & 722.64/689\\
       3201750105 & 3.98$_{-0.28}^{+0.03}$  &  1.76$_{-0.25}^{+0.01}$ &  0.99$_{-0.08}^{+0.02}$ & 6.2$_{-0.3}^{+5.1}$ & 2.82$_{-0.09}^{+0.10}$ & 714.74/730\\
       3201750106 & 4.22$_{-0.27}^{+0.22}$  &  1.82$_{-0.22}^{+0.25}$ &  0.98$_{-0.12}^{+0.33}$ & 5.1$_{-3.8}^{+6.1}$ & 3.20$_{-0.21}^{+0.20}$ & 534.67/564\\
       3201750107 & 3.90$_{-0.23}^{+0.21}$  &  1.68$_{-0.20}^{+0.19}$ &  1.01$_{-0.07}^{+0.13}$ & 7.3$_{-3.3}^{+4.1}$ & 3.13$_{-0.15}^{+0.12}$ & 685.95/620\\
       3201750108 & 3.93$_{-0.22}^{+0.21}$  &  1.68$_{-0.19}^{+0.19}$ &  1.00$_{-0.07}^{+0.12}$ & 8.3$_{-3.7}^{+4.3}$ & 3.32$_{-0.15}^{+0.13}$ & 622.64/614\\
       3201750109 & 3.80$_{-0.23}^{+0.20}$  &  1.62$_{-0.22}^{+0.20}$ &  1.11$_{-0.09}^{+0.14}$ & 8.3$_{-3.4}^{+4.8}$ & 4.61$_{-0.15}^{+0.17}$ & 668.66/633\\
       3201750110 & 4.04$_{-0.46}^{+0.07}$  &  1.90$_{-0.50}^{+0.07}$ &  1.33$_{-0.34}^{+0.08}$ & 3.5$_{-0.6}^{+10.6}$ & 4.80$_{-0.19}^{+0.22}$ & 609.36/562\\
       3201750111 & 3.88$_{-0.10}^{+0.09}$  &  1.68$_{-0.10}^{+0.10}$ &  1.11$_{-0.07}^{+0.09}$ & 6.5$_{-2.0}^{+2.4}$ & 4.95$_{-0.09}^{+0.09}$ & 925.54/810\\
       3201750112 & 3.95$_{-0.35}^{+0.10}$  &  1.91$_{-0.41}^{+0.13}$ &  1.37$_{-0.32}^{+0.12}$ & 3.7$_{-0.8}^{+7.1}$ & 4.97$_{-0.20}^{+0.22}$ & 557.71/571\\
       3201750113 & 3.87$_{-0.12}^{+0.11}$  &  1.73$_{-0.13}^{+0.28}$ &  1.18$_{-0.14}^{+0.44}$ & 4.0$_{-2.7}^{+3.2}$ & 5.27$_{-0.14}^{+0.14}$ & 719.52/740\\
       3201750114 & 3.77$_{-0.33}^{+0.28}$  &  1.51$_{-0.31}^{+0.25}$ & 0.96$_{-0.09}^{+0.19}$ & 11.0$_{-7.2}^{+9.4}$ & 4.84$_{-0.25}^{+0.41}$ & 518.81/526\\
       3201750115 & 3.79$_{-0.18}^{+0.16}$  &  1.68$_{-0.18}^{+0.23}$ &  1.11$_{-0.12}^{+0.28}$ & 5.7$_{-3.4}^{+4.9}$ & 4.88$_{-0.17}^{+0.17}$ & 731.37/672\\
       3201750116 & 3.80$_{-0.25}^{+0.25}$  &  2.12$_{-0.48}^{+0.33}$ &  2.10$_{-1.19}^{+1.67}$ & 0.4$_{-0.4}^{+0.7}$ & 5.47$_{-0.35}^{+0.18}$ & 609.95/594\\
       3201750118 & 3.68$_{-1.04}^{+0.32}$  &  2.14$_{-0.49}^{+0.53}$ &  1.66$_{-0.97}^{+3.77}$ & 0.7$_{-0.7}^{+8.8}$ & 3.46$_{-0.29}^{+0.41}$ & 421.87/394\\
       3201750119 & 3.66$_{-0.59}^{+0.67}$  &  2.32$_{-0.58}^{+0.90}$ &  1.70$_{-0.87}^{+1.66}$ & 0.9$_{-0.9}^{+1.8}$ & 3.74$_{-0.35}^{+0.44}$ & 275.17/285\\
       3201750120 & 3.48$_{-0.50}^{+0.10}$  &  2.05$_{-0.54}^{+1.89}$ &  0.95$_{-0.95}^{+2.74}$ & 1.6$_{-1.6}^{+3.1}$ & 4.18$_{-0.57}^{+0.91}$ & 181.73/150\\
       3201750126 & 3.32$_{-0.42}^{+0.52}$  &  1.24$_{-0.84}^{+0.53}$ &  0.80$_{-0.07}^{+0.07}$ & 12.2$_{-7.0}^{+5.1}$ & 1.32$_{-0.10}^{+0.14}$ & 347.31/311\\
       3201750132 & 3.73$_{-0.87}^{+0.80}$  &  1.47$_{-1.47}^{+0.81}$ &  0.80$_{-0.15}^{+0.38}$ & 10.0$_{-9.3}^{+9.0}$ & 1.33$_{-0.07}^{+0.07}$ & 127.44/106\\
       3201750133 & 3.97$_{-0.54}^{+0.61}$  &  1.56$_{-0.59}^{+0.44}$ &  0.69$_{-0.13}^{+0.09}$ & 13.4$_{-10.2}^{+9.7}$ & 1.20$_{-0.14}^{+0.22}$ & 220.25/246\\
		\hline
	\end{tabular}
\end{table*}

\begin{figure}
 \includegraphics[width=\columnwidth]{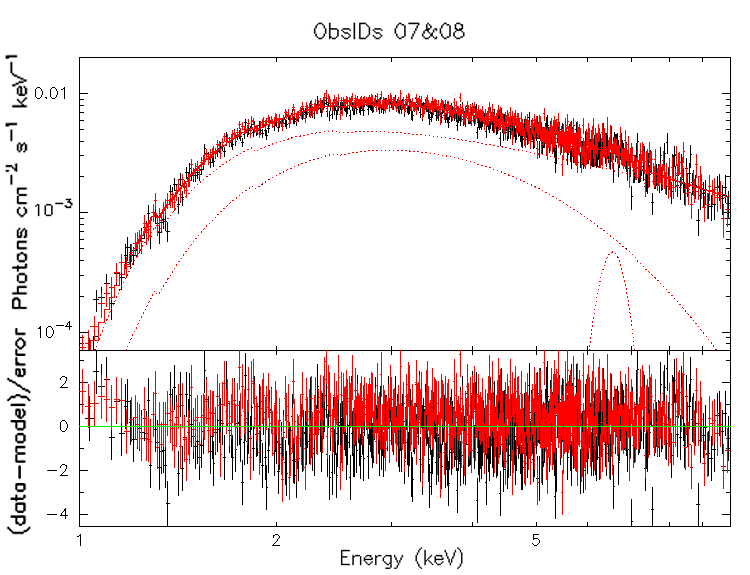}
 \caption{Unfolded spectrum of the OBSIDs 07 (black) and 08 (red) fitted simultaneously with model 2 and a gaussian component, showing the narrowest iron line component we could identify, with $\sigma \sim 0.2 keV$.}
 \label{fig:ironline}
\end{figure}

The search for the iron line in the fits with models 1 and 2 revealed the presence of similar features but often in correspondence of different OBSIDs. With the exception of a few cases, the iron line appears broad, with $\sigma \sim$ 0.6~keV and energies in the range 6.4--6.7~keV, but the parameters are loosely constrained. In order to constrain better the parameters, we simultaneously fitted the contiguous observations (see Sec. \ref{sec:continuum}). We observe an improvement in the determination of the bounds of the equivalent width with respect to the single observations. In Table \ref{tab:Fe_sum}, we report the best fit values and equivalent widths in the case of the two analysed models. When the component did not bring a significant improvement to the fit, we reported the upper limits on the equivalent width calculated with the values of energy and line width reported on Table \ref{tab:Fe_sum}. Figure \ref{fig:ironline} shows the unfolded spectra in the case of the simultaneous fitting of observations 07 and 08, which gives the narrowest iron line component and better constrained parameters.

\begin{table*}
    \centering
    \caption{Best fit parameters of the iron line component with simultaneous fitting of spectrally similar observations. The asterisks denote the group of observations where the addition was not significant and the equivalent width is reported with upper limits. Errors are at the 90\% confidence level. }
    \label{tab:Fe_sum}
\renewcommand{\arraystretch}{1.3}
    \begin{tabular}{lccccccr}
        \hline
	& \multicolumn{3}{c}{Model 1} & \multicolumn{3}{c}{Model 2} \\
	\cmidrule(r){2-4}\cmidrule(l){5-7}
        OBSIDs & Energy [keV] & $\sigma$ [keV] & EW [eV] & Energy [keV] & $\sigma$ [keV] & EW [eV] & Total counts [cts]\\
        \hline
        01--06 & <6.63 & >0.52 & 96$^{+57}_{-58}$ & <6.53 & >0.54 & 199$^{+124}_{-72}$ & 472405\\
        07--08 & 6.55$^{+0.11}_{-0.11}$ & 0.23$^{+0.12}_{-0.08}$ & 82$^{+54}_{-62}$ & 6.55$^{+0.11}_{-0.10}$ & 0.27$^{+0.24}_{-0.09}$ & 109$^{+74}_{-45}$ & 110402 \\
        09--16 & <6.59& >0.52 & 159$^{+58}_{-41}$ & <6.55 & >0.58 & 199$^{+79}_{-49}$ & 867699\\
        18--20* & 6.41 (fixed) & 0.60 (fixed) & <266 & 6.41 (fixed) & 0.60 (fixed) & <1684 & 32399\\
         26--32--33* & 6.41 (fixed) & 0.60 (fixed) & <570 & 6.77 (fixed) & 0.60 (fixed) & <1905 & 26436\\
        \hline
    \end{tabular}
\end{table*}

\subsection{Type-I X-ray burst}
\begin{figure}
 \includegraphics[width=\columnwidth]{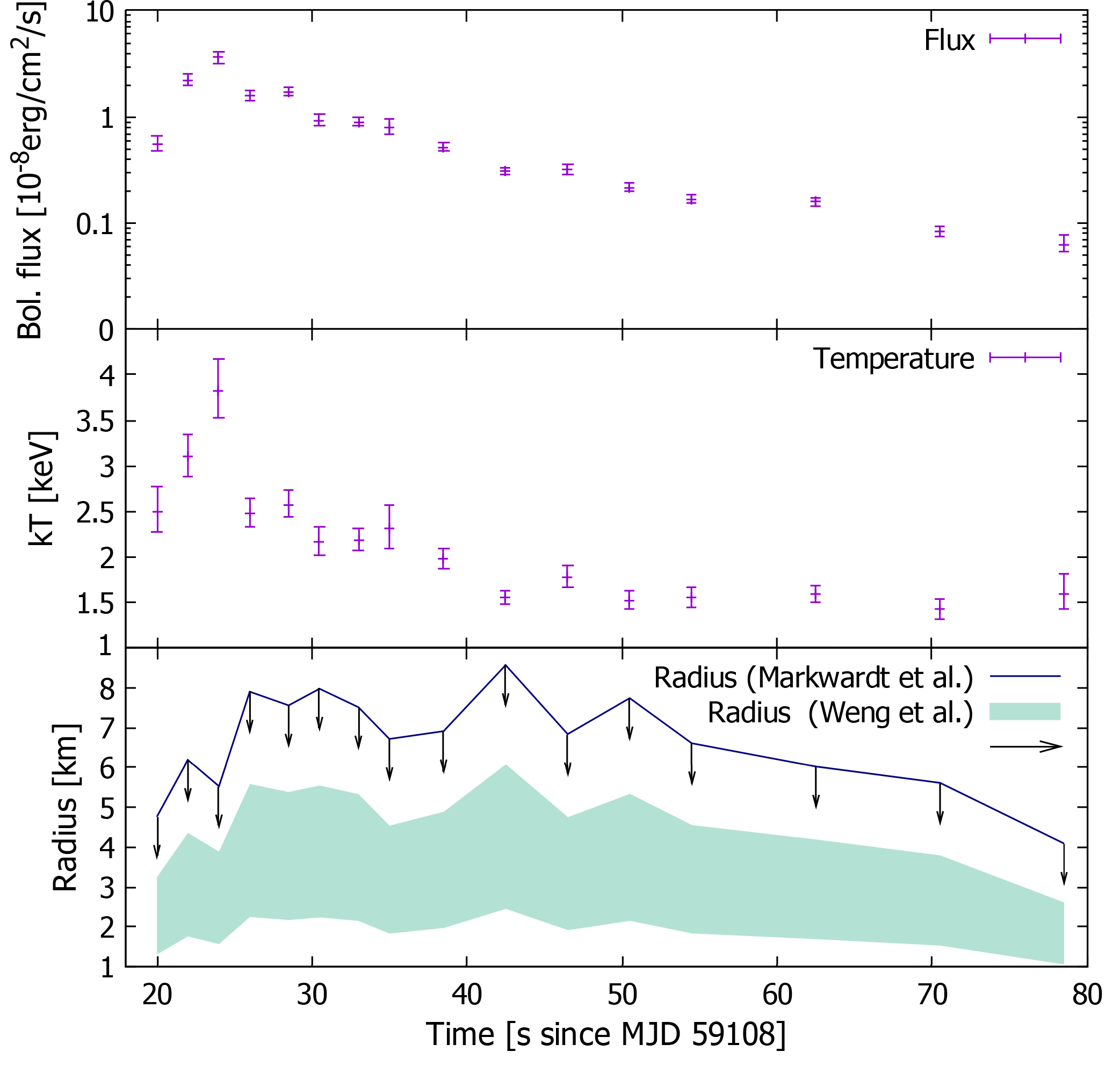}
 \caption{Temporal evolution of the parameters of the type-I X-ray burst: the green series is the unabsorbed bolometric flux; the violet series is the black body temperature. The navy blue line is the upper limit on the black body radius as derived from the upper limit on the distance by \citet{Tel1443}. The light blue band is the region of possible values derived from the interval of distances calculated by \citet{weng15}. All times are referred to MJD 59108. The errors are at a 68\% confidence level.}
 \label{fig:burst_par}
\end{figure}

The time-resolved spectral analysis did not reveal the typical behaviour of photospheric radius expansion type-I bursts. Figure \ref{fig:burst_par} shows the temporal evolution of the unabsorbed bolometric (0.1--30~keV) flux and the black body temperature. The coloured area of the third panel identifies the region of possible values for the radius, as derived from the interval of distances measured by \citet{weng15}. The blue line instead identifies the upper limit on the radius as derived from the upper limit on the distance calculated by \citet{Tel1443}. We do not observe the local minimum in the temperature and the following peak, sign of the touchdown of the NS atmosphere after the expansion. The peak temperature is reached at the beginning, equal to 3.4$^{+0.3}_{-0.2}$~keV (errors at 68\% c.l.). It must be considered that the best fit values obtained for the temperature (and consequently for the radius) from the spectral fitting return values that are larger with respect to the real effective temperature at the NS surface, with a factor depending on the chemical composition and luminosity \citep[see][for details]{Suleimanov2011}. Table \ref{tab:burst} reports the best fit values for the temporally-resolved spectroscopy of the burst. The type-I burst lasts $\sim$~80~s, typical duration of H/He bursts \citep{Galloway2021}. The peak flux is reached in the third segment of the time-resolved spectra, at a value of 3.3$^{+0.4}_{-0.3}\cdot10^{-8}$~erg~cm$^{-2}$~s$^{-1}$, which corresponds to a fraction of the flux at the Eddington limit in the range 12.3--76\%, calculated from the Eddington luminosity of 3.8$\cdot10^{38}$~erg~s$^{-1}$ \citep{Kuulkers2003} and for a mass of $1.4 M_\odot$, for the interval of distances of 3.5--8.7~kpc from \citet{weng15}. The best fit values for the temporally resolved spectral analysis are reported in Table \ref{tab:burst}.

\begin{table}
    \centering
    \caption{Best fit values of the temporally resolved spectra of the type-I burst.The bolometric flux is calculated in the 0.1--30 keV energy range. Errors are at the 68\% confidence level. }
    \label{tab:burst}
\renewcommand{\arraystretch}{1.3}
    \begin{tabular}{cccr}
        \hline
	\thead{Time since \\MJD 59108 [s]} & kT$_{BB}$ [keV] & \thead{Unabsorbed bolometric \\flux [$10^9$~erg~cm$^{-2}$~s$^{-1}$]} & Norm$_{BB}$ \\
        \hline
	4 & 2.08$^{+0.13}_{-0.15}$ &  5.29$^{+0.65}_{-0.55}$ &  26.8$^{+4.6}_{-4.0}$ \\
	 6 & 2.84$^{+0.19}_{-0.17}$ &  21.33$^{+2.44}_{-2.14}$ &  32.8$^{+4.5}_{-4.1}$ \\
	8 & 3.43$^{+0.26}_{-0.22}$ &  33.24$^{+3.99}_{-3.49}$ &  26.5$^{+3.6}_{-3.4}$ \\
	10 & 2.33$^{+0.14}_{-0.12}$ &  15.96$^{+1.64}_{-1.45}$ &  51.7$^{+6.9}_{-6.2}$ \\
	12.5 & 2.41$^{+0.13}_{-0.11}$ &  17.07$^{+1.56}_{-1.40}$ &  48.2$^{+5.1}_{-5.6}$ \\
	14.5 & 2.02$^{+0.13}_{-0.12}$ &  9.52$^{+1.06}_{-0.91}$ &  54.0$^{+8.5}_{-7.6}$ \\
	17 & 2.06$^{+0.10}_{-0.09}$ &  9.43$^{+0.78}_{-0.71}$ &  49.0$^{+5.8}_{-5.4}$  \\
	19 & 2.06$^{+0.18}_{-0.15}$ &  7.84$^{+1.19}_{-0.96}$ &  40.9$^{+8.3}_{-7.2}$ \\
	22.5 & 1.86$^{+0.08}_{-0.08}$ &  5.71$^{+0.43}_{-0.40}$ &  44.6$^{+5.3}_{-4.8}$ \\
	26.5 & 1.51$^{+0.05}_{-0.05}$ &  3.72$^{+0.21}_{-0.20}$ &  66.0$^{+7.2}_{-6.5}$ \\
	30.5 & 1.65$^{+0.08}_{-0.08}$ &  3.74$^{+0.30}_{-0.27}$ &  46.9$^{+6.6}_{-5.9}$ \\
	34.5 & 1.45$^{+0.07}_{-0.07}$ &  2.75$^{+0.20}_{-0.18}$ &  58.2$^{+8.6}_{-7.6}$ \\
	38.5 & 1.49$^{+0.07}_{-0.07}$ &  2.35$^{+0.17}_{-0.16}$ &  44.6$^{+6.4}_{-5.7}$ \\
	46.5 & 1.47$^{+0.06}_{-0.05}$ &  2.15$^{+0.12}_{-0.12}$ &  42.9$^{+4.9}_{-4.5}$ \\
	54.5 & 1.37$^{+0.06}_{-0.06}$ &  1.47$^{+0.09}_{-0.08}$ &  38.2$^{+5.0}_{-4.5}$ \\
	62.5 & 1.42$^{+0.08}_{-0.07}$ &  1.23$^{+0.09}_{-0.08}$ &  28.3$^{+4.6}_{-4.0}$ \\
        \hline
    \end{tabular}
\end{table}

\subsection{Timing results}
\begin{figure}
 \includegraphics[width=\columnwidth]{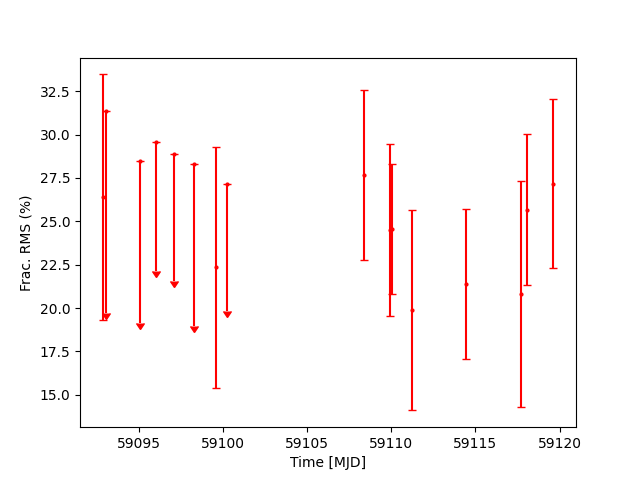}
 \caption{Evolution in time of the fractional RMS in the $0.1-100$~Hz frequency range, expressed in percentage, throughout the outburst. The error bars and upper limits are at 3$\sigma$ level.}
 \label{fig:rms}
\end{figure}

The PDS for each observation do not show strong peaked features, such as quasi-periodic oscillations, but are characterised by a red noise in the entire range of observed frequencies. Figure \ref{fig:rms} shows the temporal evolution of the fractional RMS in the $0.1-100$~Hz frequency range, reported also in the Table \ref{tab:rms}. The fractional RMS varies between $\sim 20 \%$ and $\sim 30 \%$, with some observations giving only upper limits. In these cases, the absence of a strong signal in the PDS does not allow for a better estimation of the fractional RMS.

\begin{table}
    \centering
    \caption{Fractional RMS values calculated in the 0.1--100~Hz frequency range. Errors and upper limits are at the 3$\sigma$ confidence level.}
    \label{tab:rms}
\renewcommand{\arraystretch}{1.3}
    \begin{tabular}{cc}
        \hline
	Time [MJD] & Frac. RMS [\%] \\
        \hline
	59092.85 & 26.4 $\pm$ 7.1\\
	59093.04 & <31.4 \\
	59095.05 & <28.5\\
	59096.01 & <29.6\\
	59097.11 & <28.9\\
	59098.28 & <28.3\\
	59099.57 & 18.9 $\pm$ 9.4\\
	59100.22 & <27.2\\
	59108.41 & 27.7 $\pm$ 4.9\\
	59109.95 & 24.5 $\pm$ 5.0\\
	59110.03 & 24.6 $\pm$ 3.8\\
	59111.25 & 19.9 $\pm$ 5.8\\
	59114.47 & 21.4 $\pm$ 4.3\\
	59117.70 & 20.8 $\pm$ 6.5\\
	59118.09 & 25.7 $\pm$ 4.4\\
	59119.64 & 27.2 $\pm$ 4.9\\
        \hline
    \end{tabular}
\end{table}

\section{Discussion}
\subsection{The continuum model}
XTE J1810--189 is a NS LMXB system that, during the 2020 outburst, reached luminosities of the order of 10$^{36}$~erg~s$^{-1}$. We identified three different models that fit the data well: model 1, a single Comptonised component; model 2, a cold, seed photons Comptonisation and a black body; model 3, a black body plus a power-law component.

Our analysis revealed that all 22 observations of XTE J1810--189 collected by \textit{NICER} can be well fit by an absorbed, thermally Comptonised component (model 1) whose photon index slightly varies in time, reaching its highest value towards the end of the outburst. We could not identify the direct black body component, which must provide the seed photons required to generate the Comptonised component. The temperature of the Comptonised black body is found to be $\sim$~0.6~keV, which can either be a hot disc or the NS surface.

In his overview, \citet{Barret2001} justifies the absence of a direct emission component in the hard states of LMXB with three main reasons: the direct component is absent, too faint or outside the observed energy range. In order to test whether the component is present but obscured, we derived the optical depth of the Comptonising region. The energy range in which the study was conducted does not allow to determine a best fit value of the electron temperature of the Comptonising region, therefore we have to base our reasoning on an upper limit for the optical depth of the Comptonising corona. The values reported in the last column of Table \ref{tab:model1} are compatible with the presence of a thick region, with at least 2\% of detectable direct emission. The values in the table represent an upper limit, but even for kT$_e$ values of 30~keV, we get a moderately thick corona with $\tau$ between 1.7 and 2 ($\sim$15\% of detectable direct emission). It is therefore possible that our difficulty in determining a significant direct component derives from the presence of an absorbing region.

A similar situation has been reported by \citet{Zand1999} in the case of the NS LMXB SAX J1748.9--2021 during the 1998 outburst, in which it was found to be in a hard, low luminosity state. The source could be well fitted by both models composed by a black body component and a broken power-law (similar to our model 3), and a single Comptonised component, with no direct emission detected. Their energy range was larger and they could properly constrain the electron temperature and derive the optical depth of the Comptonising region, which was $\sim 6$ for a spherical geometry (as in our case) and $\sim 2.7$ for a disc geometry.

We tested a more sophisticated model where we assume the Comptonised seed photons come from the accretion disc, but its temperature is too cold to be directly observed (fixed to 0.1~keV). The additional black body component represents the direct emission from the neutron star, as confirmed by the estimated size of the black body emitting region, i.e. a few km, compatible with a hot spot on the NS surface. Figure \ref{fig:parameters} shows a similar temporal evolution of the seed-photon temperature of the Comptonised component of model 1 and the direct component of model 2 (which we hypothesise to be the neutron star).

The detection of an emission line at $\sim 6.4$~keV can be linked to the Fe K-shell emission line, arising from the fluorescence of iron in the inner parts of the accretion disc. It has been detected in several single observations with both models 1 and 2, but not always in the same OBSIDs. We grouped the observations according to their spectral properties and the results are similar between the models 1 and 2. The addition of the iron line component was significant for the observations at the beginning of the outburst, but we could only set limits on the parameters, with $\sigma$ larger than 0.5~keV. The combined observations 07 and 08 show a much narrower line, which derives mostly from the contribution of observation 07 where it is significantly detected in the single observation. In this case, we were able to properly constrain the parameters and we identified a component with $\sigma \sim 0.25$~keV. The equivalent width is loosely constrained, and for the final observations, the iron line component was found to be not significant. The final two groups of observations are also the ones with lower statistics, which can be a cause for the non detection.

\subsection{Spectral behaviour}
If we consider the previous estimate of the upper limit for the distance of the system of 11.5~kpc \citep{Tel1443}, we obtain an upper limit on the luminosity of $\sim 10^{36}$~erg~s$^{-1}$, similar to what was obtained in some Very-Faint X-ray Transients \citep[VFXTs,][]{Wijnands2015}. These systems are LMXBs in which the X-ray luminosity in the 0.5--10~keV energy band ranges between $10^{34}$--$10^{36}$~erg~s$^{-1}$, and they show a softening of the spectra (higher photon index) with decreasing luminosity. It is still not clear whether the softening of the spectra arises from a soft component (the NS surface or boundary layer) becoming more important due to low-level accretion, or from the evolution of the Comptonised component. This behaviour has been reported before for XTE J1810--189 \citep{weng15} and even in this work, we can notice that spectra have a higher photon index towards the end of the outburst, when the flux is decreasing. XTE~J1810--189 fits into the category of Hard X-ray Transients, LMXBs that remain into a hard state throughout their outburst. The reason of such spectral evolution is still not completely understood. It could either be an accretion disc truncated by the magnetosphere at large distances or an intrinsic property of the mass accretion rate.

X-ray faint and rather hard outbursts are also observed in Accreting Millisecond X-ray Pulsars (AMXP), where the luminosity typically does not exceed 10\% of the Eddington luminosity in the outburst. The similar behaviour of the two classes of objects was compared by \citet{Wijnands2015} in order to verify a possible common nature, but no final conclusion could be drawn. AMXPs spectra are composed by a Comptonised component and one (or two) black body components, with electron temperatures around tens of keV \citep{DiSalvo2020}. There can also be iron line components, broadened by relativistic effects, and reflection components. The presence of these features is not always observed, though \citep[see][for details and more examples]{DiSalvo2020, Sanna2018_16597, Sanna2018_17379}. The spectral behaviour is similar to what we observed in XTE J1810--189, however no pulsations have ever been detected.

The duration of the type-I burst can give us some information regarding the companion star of the system. The mixed H/He composition hint at a hydrogen-rich main sequence star. Nevertheless, there are sources, such as Ultra-Compact X-ray Binaries \citep[UCXBs; see, for a recent review,][]{Padilla2023}, that show bursts of similar duration, along with shorter, He-fueled bursts, although primarily fueled by H-poor material \citep{Zand2007}. These systems are characterised by orbital periods of P$_{orb} \approx 1$~h, meaning that the donor star must be an evolved, hydrogen-poor and helium-rich star. In fact, the small dimensions of the orbit imply the presence of a small disc. UCXBs are characterised by low luminosities and several of them are also VFTXs. Type-I bursts in UCXBs are also likely to reach the Eddington limit and become photospheric radius expansion due to the fact that the 3$\alpha$ processes for the burning of helium are very rapid. LMXBs systems can be identified as UCXBs if their persistent luminosity is $\sim 1 \%$ L$_{Edd}$, condition that our source satisfies. No final conclusion can be done without a measurement of the orbital period of the system or without direct information on the composition of the companion star, since the duration of the type-I burst does not offer an univocal answer.

\subsection{Timing analysis}
The PDS show a red noise component that is compatible with Poisson noise. The values of the fractional rms are typical of systems in the low/hard state \citep{MunozDarias2014}, in which the Comptonising region is contributing the most to the power. This is in line with the results of the spectral analysis that imply a predominance of the Comptonised component. The rms does not seem to vary appreciably during the outburst, even if our measures are affected by large uncertainties. These results are in favour of the VFTX interpretation, as these systems typically show rms values $\sim 20 \%$  \citep[see, for example,][]{ArmasPadilla2017}.

\subsection{The geometry of the system}
To our two possible models correspond two possible system geometries. In the case of model 1, we have a moderately thick region that obscures the central NS and part of the inner regions of the accretion disc. The outer regions are not directly detectable because their temperatures are outside our energy range. In the case of model 2, the disc is colder and therefore not directly detectable. We still detect an iron emission line and, even if it is unlikely for a low luminosity/hard state, we cannot exclude that, in this case, it might arise from recombination processes within the Comptonising region. The disc can be truncated by the magnetic field, or we might be in the presence of an advection-dominated accretion flow (ADAF). The radiatively inefficient process, typical of system with low accretion rates, causes the disc to swell in its innermost region. While the two models are statistically equivalent, model 1 appears to be more physically realistic.

\section{Conclusions}

In this paper, we studied the spectral evolution of the NS LMXB XTE J1810--189 during its latest outburst. We analysed \textit{NICER} data in the 1--10~keV energy range, modelling the emission with an absorbed, thermally Comptonised component, and with a different model composed by a cold, Comptonised disc and black body. The photon index ranges from $\sim$~1.7 to $\sim$~2.2. We cannot tell yet whether the seed photons for the Comptonisation component are coming from the accretion disc or the neutron star. The seed-photon temperature of the Comptonised component varies between $\sim$~0.47 and $\sim$~0.79~keV. Upper limits on the optical thickness suggest a moderately thick Comptonising region, which can hinder the detection of the direct emission from the central object. During the three months, which cover the entire outburst, we found that the source did not reach a full high/soft state and the highest luminosity was found to be $\sim 10^{36}$~erg~s$^{-1}$. We detected a broad iron K-shell emission line at $\sim 6.4$~keV in a couple of observations and likely the low statistics hindered the detection in the final observations of the outburst. We analysed the type-I X-ray burst, whose duration suggest H/He accreting material.

\section*{Acknowledgements}
The authors acknowledge financial contribution from the agreement ASI-INAF n.2017-14-H.0 and INAF mainstream (PI: A. De Rosa, T. Belloni), from the HERMES project financed by the Italian Space Agency (ASI) Agreement n. 2016/13 U.O and from the ASI-INAF Accordo Attuativo HERMES Technologic Pathfinder n. 2018-10-H.1-2020. We also acknowledge support from the European Union Horizon 2020 Research and Innovation Frame- work Programme under grant agreement HERMES-Scientific Pathfinder n. 821896 and from PRIN-INAF 2019 with the project "Probing the geometry of accretion: from theory to observations" (PI: Belloni).

A. Marino is supported by the H2020 ERC Consolidator Grant "MAGNESIA" under grant agreement No. 817661 (PI: Rea) and National Spanish grant PGC2018-095512-BI00. This work was also partially supported by the program Unidad de Excelencia Maria de Maeztu CEX2020-001058-M, and by the PHAROS COST Action (No. CA16214).
\section*{Data Availability}
 
The data of this paper have been collected by the NICER mission and are publicly available through NASA's HEASARC database: \url{https://heasarc.gsfc.nasa.gov/cgi-bin/W3Browse/w3browse.pl}.


\bibliographystyle{mnras}
\bibliography{bibliography}

\bsp	
\label{lastpage}
\end{document}